\documentclass[twocolumn,showpacs,prd]{revtex4}
\usepackage{mathrsfs}
\usepackage{longtable,lscape}
\usepackage{txfonts}
\usepackage{amssymb}
\usepackage{indentfirst}
\usepackage{graphicx,,booktabs}
\usepackage{color}
\usepackage{amssymb}

\begin{document}

\title{Dipion invariant mass distribution of the anomalous
$\Upsilon(1S) \pi^{+} \pi^{-}$ and $\Upsilon(2S) \pi^{+} \pi^{-}$
production near the peak of $\Upsilon(10860)$}
\author{Dian-Yong Chen$^{1,3}$}
\author{Jun He$^{1,2}$}\author{Xue-Qian Li$^4$}
\author{Xiang Liu$^{1}$\footnote{Corresponding author}
}\email{xiangliu@lzu.edu.cn}
\affiliation{
$^1$Research Center for Hadron and CSR Physics,
Lanzhou University and Institute of Modern Physics of CAS, Lanzhou 730000, China\\
$^2$School of Physical Science and Technology, Lanzhou University, Lanzhou 730000,  China\\
$^3$Nuclear Theory Group, Institute of Modern Physics of CAS, Lanzhou 730000, China\\
$^4$Physics department, Nankai University, Tianjin 300071, China}
\date{\today}

\begin{abstract}
Considering the defects of the previous work for estimating the anomalous production rates of
$e^+e^-\to \Upsilon(1S)\pi^+\pi^-,\,\Upsilon(2S)\pi^+\pi^-$ near the peak of the $\Upsilon(5S)$ resonance at $\sqrt s=10.87$ GeV [K.F. Chen {\it et al}. (Belle Collaboration), Phys. Rev. Lett. {\bf 100}, 112001 (2008)],
we suggest a new scenario where the contributions from the direct dipion transition and the final state interactions interfere to
result in not only the anomalously large production rates, but also the lineshapes of the differential widths consistent with the experimental measurement when assuming
the reactions are due to the dipion emission of $\Upsilon(5S)$. At the end, we raise a new puzzle that the predicted differential width
$d\Gamma(\Upsilon(5S)\to\Upsilon(2S)\pi^+\pi^-)/d\cos\theta$ has a discrepant trend from the data while other predictions are well in accord with the
data. It should be further clarified by more accurate measurements carried by future experiments.

\end{abstract}

 \maketitle

Looking at the spectra of the $b\bar b$ system listed in the particle data book \cite{Nakamura:2010zzi}, there exist six bottomonia with $J^{PC}=1^{--}$, which are, respectively,
$\Upsilon(nS)$ (n=1,2,3,4), $\Upsilon(10860)$, and $\Upsilon(11020)$. The first five resonances are orderly assigned as $nS$ ($n=1,\cdots,5$) $b\bar{b}$ states,
whereas the extra one, $\Upsilon(11020)$, may be the $6S$ state. The estimate on their spectra in the potential model supports such assignments \cite{Godfrey:1985xj,Brambilla:2004wf,Brambilla:2010cs}. However, recently anomalous large
rates of $e^+e^-\to \Upsilon(1S)\pi^+\pi^-,\,\Upsilon(2S)\pi^+\pi^-$ near the peak of the $\Upsilon(5S)$ resonance at $\sqrt s=10.87$ GeV were observed by
the Belle Collaboration to be
larger than the dipion-transition rates between the lower members of the $\Upsilon$ family by 2 orders of magnitude.
The Belle data are $\Gamma(\Upsilon(5S)\to
\Upsilon(1S)\pi^+\pi^-)=0.59\pm0.04(\mathrm{stat})\pm0.09(\mathrm{syst})$
MeV and $\Gamma(\Upsilon(5S)\to
\Upsilon(2S)\pi^+\pi^-)=0.85\pm0.07(\mathrm{stat})\pm0.16(\mathrm{syst})$
MeV \cite{Abe:2007tk}.
The Belle observation has stimulated theorists' extensive interest in exploring the reason that results in
such anomalous phenomena.

There are two possibilities that may offer reasonable interpretations of the anomalous large rates. First, these anomalous production rates announced by Belle are from an exotic resonance structure different from
$\Upsilon(10860)$. The second is that there may exist extra contributions that differ from the direct dipion emission
$\Upsilon(10860) \to \Upsilon(1S,2S)\pi^+\pi^-$. It is expected that with a careful analysis we may eventually identify the reasonable, or at least the dominant
source of the large rates. Thus one not only obtains the branching ratios, but also needs to fit the lineshapes of the differential widths over the invariant mass of
dipion and over the angular distribution $\cos\theta$.

Along the first route, Ali {\it et al.} suggested  a tetraquark
interpretation of $Y_b(10890)=[bq][\bar{b}\bar{q}]$ \cite{Ali:2009pi,Ali:2009es,Ali:2010pq} and
analyzed the Belle data \cite{Abe:2007tk} for the anomalous
$\Upsilon(1S) \pi^{+} \pi^{-}$ and $\Upsilon(2S) \pi^{+} \pi^{-}$
productions near the $\Upsilon(5S)$ resonance. By fitting the ${\pi^+
\pi^-}$ invariant mass spectrum and the $\cos \theta$ distributions
for $Y_{b}(10890) \to \Upsilon(nS) \pi^{+} \pi^{-}$ $(n=1,2)$ shown in
Figs. 1 and 2 of Ref. \cite{Ali:2009es}, they claimed that the
tetraquark interpretation can well describe the anomalous rates of the
two-pion-production. In their scenario, there are non-resonant and resonant
contributions interfering to result in the branching ratio and differential widths.
In that work \cite{Ali:2009es,Ali:2010pq}, a simple Lorentz structure Lagrangin
was introduced to stand as the effective interaction.

An alternative scenario was proposed, namely, the final state interaction of
$\Upsilon(10860)$ decaying into $\Upsilon(1S,2S)\pi^+\pi^-$ can be realized via
sub-processes $\Upsilon(10860)\to B^{(*)}\bar B^{(*)}\to \Upsilon(1S,2S)\pi^+\pi^-$  \cite{Meng:2007tk,Meng:2008dd,Simonov:2008ci}.
It was also claimed that the anomalous $\Upsilon(1S)
\pi^{+} \pi^{-}$ and $\Upsilon(2S) \pi^{+} \pi^{-}$ productions near
the $\Upsilon(5S)$ resonance receive reasonable explanations.
The tetraquark interpretation presented in Ref. \cite{Ali:2009es} is evidently not unique,
and, moreover below
we will show that the dipion invariant mass distribution and the angular
distribution of $Y_b(10890)\to \Upsilon(2S)\pi^+\pi^-$ obtained with the teraquark picture
proposed in Ref. \cite{Ali:2009es} cannot explain the Belle data well.

\begin{figure}[htb]
\begin{center}
\includegraphics[bb=600 390 20 520,scale=0.70,angle=0]{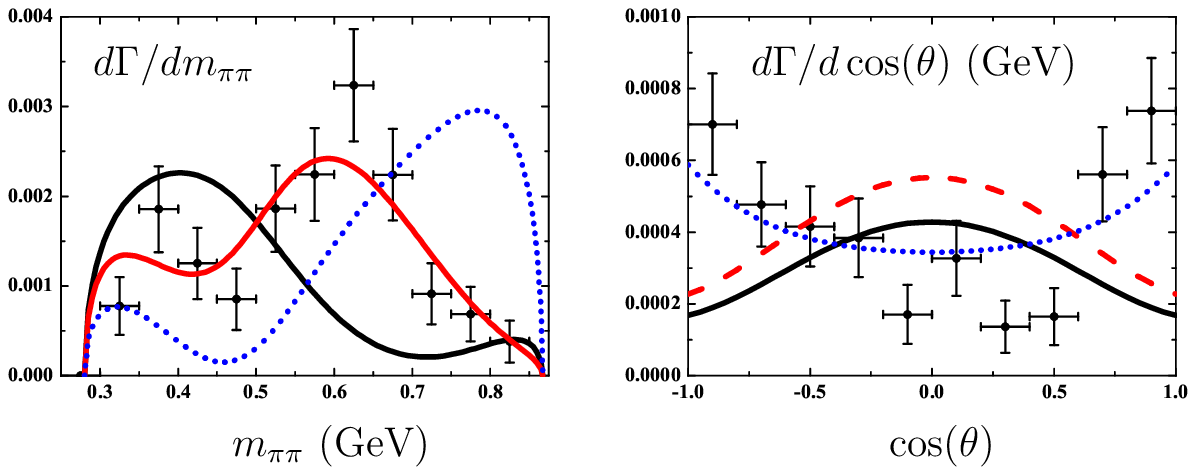}
\caption{(color online). Dipion invariant mass ($m_{\pi\pi}$)
distribution (left-hand side) and the $\cos \theta$ distribution
(right-hand side) of $\Upsilon(2S) \pi^{+}\pi^{-}$ production. The dots
with error bars are the results measured by Belle. The solid lines
denote the results reproduced with the parameters shown in Table.
II of Ref. \cite{Ali:2009es}. With the same formula as that in Ref.
\cite{Ali:2009es}, we refit the experimental data. Here, the red
dashed-line curves and blue dotted-line curves are the fitting results with
parameters \{$F=0.933\pm0.396$, $\beta=0.692\pm0.202$,
$f_\sigma=9.405\pm2.409$, $\phi=-0.460\pm0.245$ Rad\} and
\{$F=1.056\pm1.348$, $\beta=0.467\pm0.578$, $f_\sigma=10.354\pm
15.163$, $\phi=-1.785\pm1.223$ Rad\}, respectively, which correspond
to the best fits to the dipion invariant mass distribution and the
$\cos\theta$ distribution, respectively. \label{Fig-Fit-ali-1}}
\end{center}
\end{figure}

Using the formulas and parameters given in Ref. \cite{Ali:2009es} to fit the data
given by the Belle Collaboration, we find obvious discrepancies. Namely, as shown in the following figure,
employing the formulas and parameters given in Ref. \cite{Ali:2009es}, we
obtain the solid lines for the dipion invariant mass distribution and angular distribution for $Y_b(10890)\to \Upsilon(2S)\pi^+\pi^-$ (see Fig. \ref{Fig-Fit-ali-1}), and obviously
they do not fit the data points which are marked in the figure. If, with the formulas given by the authors of Ref. \cite{Ali:2009es},
we fit the dipion invariant mass distribution (the dashed line)
as the left-hand diagram of Fig. \ref{Fig-Fit-ali-1} to gain the model parameters, then applying the parameters to calculate the angular distribution $d\Gamma/d\cos\theta$,
we would have the dashed line on the right-side of Fig. \ref{Fig-Fit-ali-1}. Inversely, if we first fit the differential width $d\Gamma/d\cos\theta$ to fix the parameters
(the dotted-line on the right-hand side of Fig. \ref{Fig-Fit-ali-1}), using those parameters would result in the dotted-line on the left-hand side diagram of Fig. \ref{Fig-Fit-ali-1}. All the results
contradict the data; even the trend does not coincide. Therefore, it seems that the tetraquark scenario does not work well to some extent.

In Ref. \cite{Meng:2007tk}, the authors considered the sub-processes
$\Upsilon(10860)\to B^{(*)}\bar B^{(*)}\to \Upsilon(1S,2S)\pi^+\pi^-$ which are supposed to be the final state interaction of $\Upsilon(10860)\to  \Upsilon(1S,2S) \pi^+\pi^-$. They concluded that as the absorptive (imaginary) part of the triangle diagrams dominate, one can expect an enhancement of $200\sim600$ times compared to the partial widths of dipion emission
$\Upsilon(nS)\to\Upsilon(1S,2S)\pi^{+}\pi^{-}$ $(n\leq 3)$. Moreover, even though for $\Upsilon(4S)$, the $B\bar B$ channel is open, but
due to the limit in phase space, the $p$-value suppresses the contribution from the $B\bar B$ intermediate states. The data show
that the partial width of $\Upsilon(4S)\to \Upsilon(1S,2S)\pi^+\pi^-$ is only 2$\sim 4$ times larger than that of
$\Upsilon(3S)\to \Upsilon(1S,2S)\pi^+\pi^-$ by the Belle and Babar measurements. Thus it seems that
the largeness of the dipion emission of $\Upsilon(10860)$ can be explained as coming from the final state interactions. However, in Ref. \cite{Meng:2007tk}, the authors did not give a fit to the line shapes of the dipion invariant mass distribution $d\Gamma/dm_{\pi^+\pi^-}$ and the differential width $d\Gamma/d\cos\theta$ explicitly, so they claimed that their results were roughly consistent with data.

Even though the partial width of $\Upsilon(10860)\to \Upsilon(1S,2S)\pi^+\pi^-$ is larger than that of
$\Upsilon(nS)\to \Upsilon(mS)\pi^+\pi^-$ $(n=3,4,\, m=1,2)$
by two orders, there is still no compelling reason to ignore
the contribution from the direct dipion transition process. Especially an interference between the direct transition and the contribution
from the final state interaction (intermediate heavy-meson loops) may result in a sizable change to each scenario.
We notice that
$\Gamma(\Upsilon(10860)\to \Upsilon(2S)\pi^+\pi^-)=0.85\pm 0.07(\mathrm{stat})\pm 0.16 (\mathrm{syst})$ MeV which is larger than
$\Gamma(\Upsilon(10860)\to \Upsilon(1S)\pi^+\pi^-)=0.59\pm 0.04(\mathrm{stat})\pm 0.09 (\mathrm{syst})$ MeV \cite{Abe:2007tk}, even though the later one has a larger final-state phase space.
If the largeness is due to the composition of $\Upsilon(10860)$, this relation should be inverted. This
intriguing experimental fact reported by Belle \cite{Abe:2007tk} seems to be evidence to support an interference between the direct transition and the contribution
from the final state interaction for $\Upsilon(10860)\to \Upsilon(1S,2S)\pi^+\pi^-$. Obviously,
the final state interaction that is realized via the heavy-meson loops is a simplified version of the complicated multichannel dynamics  \cite{Simonov:2008ci}.
The direct dipion transition was dealt with in terms of the
QCD multipole expansion method where there are two color-E1 transition and the two color fields eventually hadronize into two pions. Yan and Kuang
\cite{Kuang:1981se} established the theoretical framework for the multipole expansion method, where the intermediate state between the two E1 transitions
the quark pair $Q\bar Q$ resides in a color-octet where $Q$ stands for heavy quark $b$ or $c$.
Here, we do not intend to calculate the contribution from the direct
transition, but set it as an effective interaction with the free parameter, which will be fixed by fitting data.

In this work, we suggest that the total decay amplitude
of $\Upsilon(5S)\to \Upsilon(1S,2S)\pi^+\pi^-$ should include a few terms such as
\begin{eqnarray}
\label{model} \mathcal{M}_{\mathrm{total}}&=&\mathcal{M}
\textnormal{\Huge{$[$}}
\raisebox{-18pt}{\includegraphics[width=0.2%
\textwidth]{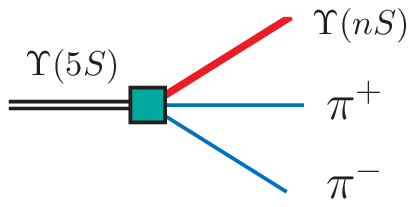}}\textnormal{\Huge{$]$}}\nonumber\\&&+\sum\limits_{R}e^{i\phi_R^{(n)}}\mathcal{M}\textnormal{\Huge{$[$}}
\raisebox{-18pt}{\includegraphics[width=0.2%
\textwidth]{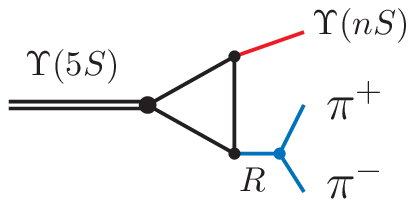}}
\textnormal{\Huge{$]$}},\label{decay}
\end{eqnarray}
where we take into account
contributions from different intermediate resonances $R$ to dipion, i.e.,
$R=\{\sigma(600),\,f_0(980),\,f_2(1270)\}$ to the $\Upsilon(1S)\pi^+\pi^-$
channel and $R=\{\sigma(600),\,f_0(980)\}$ to the $\Upsilon(2S)\pi^+\pi^-$
channel, which are allowed by the phase spaces. The phase angles
$\phi_R^{(n)}$ are introduced, which will be fixed in
this paper.

In general, the decay amplitude of the direct production of
$\Upsilon(5S)\to \Upsilon(1S,2S)\pi^+\pi^-$ is expressed as
\begin{eqnarray}
&&\mathcal{M}[\Upsilon(5S)\to
\Upsilon(nS)(p_1)\pi^+(p_2)\pi^-(p_3)]\nonumber\\&&
=\epsilon_{\Upsilon(5S)}\cdot
\epsilon_{\Upsilon(nS)}^*\Big\{\Big[q^2-\kappa(\Delta
M)^2\Big(1+\frac{2m^2_\pi}{q^2}\Big)\Big]_{\mathrm{S-wave}}\nonumber\\&&
\quad+\Big[\frac{3}{2}\kappa\big((\Delta M)^2-q^2\big)
\Big(1-\frac{4m_\pi^2}{q^2}\Big)
\Big(\cos\theta^2-\frac{1}{3}\Big)\Big]_{\mathrm{D-wave}}\Big\}
\mathcal{A},\nonumber\label{direct}
\end{eqnarray}
which was first written by Novikov and Shifman in Ref.
\cite{Novikov:1980fa} while studying $\psi^\prime\to J/\psi\pi^+\pi^-$
decay, where the S-wave and D-wave contributions are distinguished
by the subscripts S-wave and D-wave. $\Delta M$ denotes the mass
difference between $\Upsilon(5S)$ and $\Upsilon(nS)$.
$q^2=(p_2+p_3)^2\equiv m_{\pi^+\pi^-}^2$ is the invariant mass of
$\pi^+\pi^-$. $\theta$ is the angle between $\Upsilon(5S)$ and
$\pi^-$ in the $\pi^+\pi^-$ rest frame. In Ref. \cite{Ali:2009es},
Ali {\it et al.} also adopted the expression in Eq. (\ref{direct})
and introduced an extra form factor $\mathcal{A}=F/f_\pi^2$ with
$f_\pi=130$ MeV.

\begin{center}
\begin{figure}[htb]
\begin{tabular}{cccc}
\scalebox{0.45}{\includegraphics{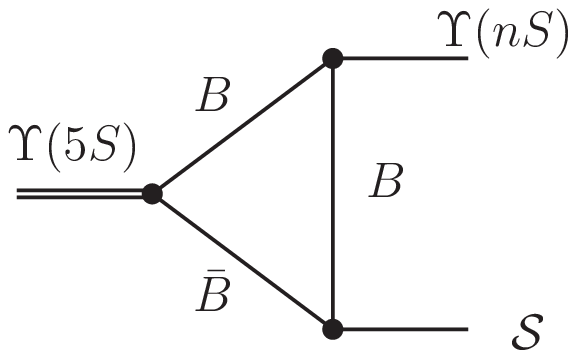}}&
\scalebox{0.45}{\includegraphics{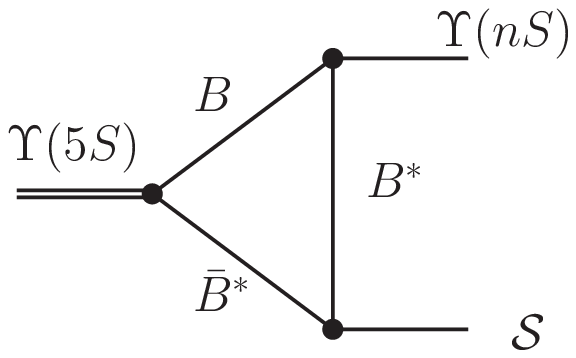}}&\scalebox{0.45}{\includegraphics{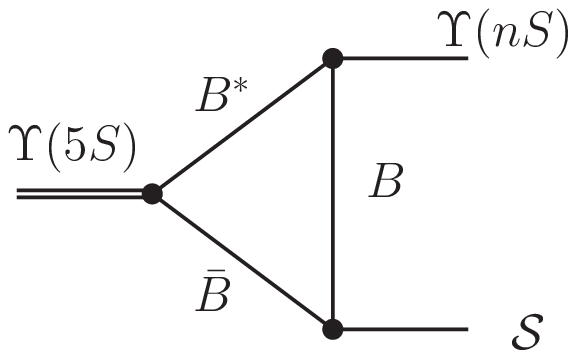}}\\
(a)&(b)&(c)\\
\scalebox{0.45}{\includegraphics{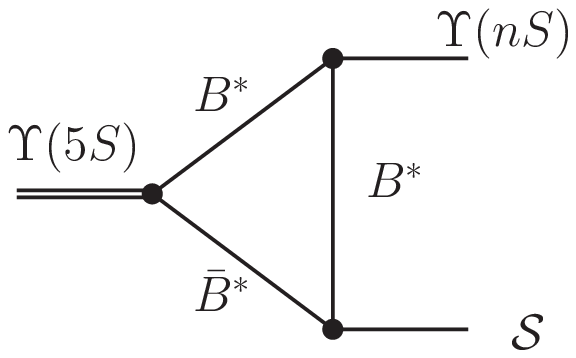}}&
\scalebox{0.45}{\includegraphics{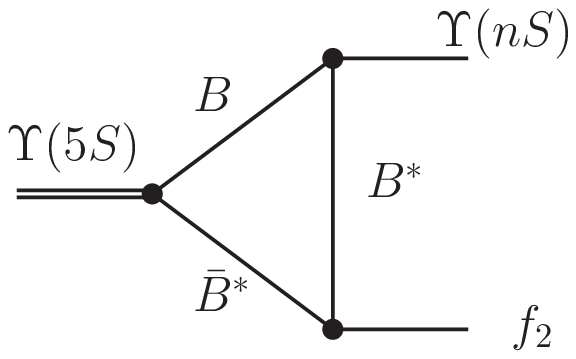}}&\scalebox{0.45}{\includegraphics{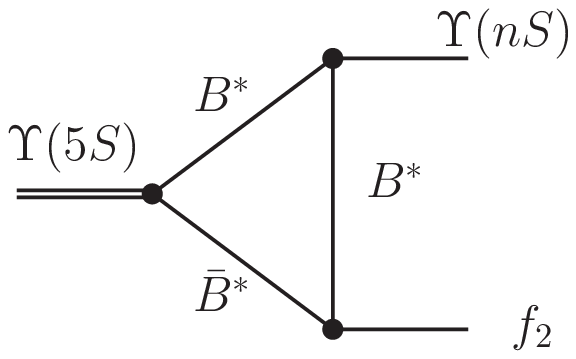}}\\
(d)&(e)&(f)
\end{tabular}
\caption{The schematic diagrams for $\Upsilon(5S)$ decays into $\Upsilon(nS)\mathcal{S}$ and $\Upsilon(nS)f_2(1270)$ $(n=1,2)$ via $B^{(*)}$ meson loops. \label{scalar}}
\end{figure}
\end{center}

As shown in Fig. \ref{scalar}, there are six diagrams corresponding to $\Upsilon(5S)$
decays into $\Upsilon(nS)\mathcal{S}$ and $\Upsilon(nS)f_2(1270)$ respectively, and $S$ and $f_0(980)$ eventually turn into
two pions; thus, as on-shell intermediate states they contribute to
$\Upsilon(5S)\to \Upsilon(nS)\pi^+\pi^-$.  Such subsequent processes are attributed to the final state interaction.
In this work, we adopt the
effective Lagrangian approach to write out the decay amplitudes for the diagrams in
Fig. \ref{scalar}, where the relevant Lagrangians include

\begin{eqnarray}
{\cal L}_{\Upsilon\cal{BB}}&=&i g_{\Upsilon \cal{BB}}\Upsilon_{\mu}(\partial^{\mu}\cal{BB}^{\dag}-\cal{B\partial^{\mu}B}^{\dag}),\\
{\cal L}_{\Upsilon\cal{B^*B}}&=&-i g_{\Upsilon \cal{B^*B}}\varepsilon^{\mu\nu\alpha\beta}\partial_{\mu}\Upsilon_{\nu}(\partial_{\alpha}\cal{B}_{\beta}^*\cal{B}^{\dag}+\cal{B\partial_{\alpha}B}^{*\dag}_{\beta}),\\
{\cal L}_{\Upsilon\cal{B^*B^*}}&=&-i g_{\Upsilon {\cal B^*B^*}}\{\Upsilon^{\mu}(\partial_{\mu}{\cal B^{*\nu}B^{*\dag}_{\nu}}-{\cal B^{*\nu}}\partial_{\mu}{\cal B^{*\dag}_{\nu}})\nonumber\\&&+
(\partial_{\mu}\gamma_{\nu}{\cal B^{*\nu}}-\Upsilon_{\nu}\partial_{\mu}{\cal B^{*\nu}})\cal{B^{*\mu\dag}}\nonumber\\
&&+{\cal B^{*\mu}}(\Upsilon^{\nu}\partial_{\mu}{\cal B}^{*\dag}_
{\nu}-\partial_{\mu}\Upsilon_{\nu}\cal{B}^{*\nu\dag})\},
\end{eqnarray}
and
\begin{eqnarray}
{\cal L}_{\cal SB^{(*)}B^{(*)}}=g_{\cal BBS}{\cal SBB^{\dag}}-g_{\cal B^*B^*S}{\cal SB^*B^{*\dag}}
\end{eqnarray}
where  ${\cal B}=(\bar{B^0},B^-,B_s^-)$ and $({\cal B}^{\dag})^{T}=(B^0,B^+,B_s^+)$. Thus, the decay amplitudes corresponding to Figs. \ref{scalar}(a)-\ref{scalar}(f) are expressed as
\begin{eqnarray}
\nonumber\mathcal{M}_a&=&(i)^3\int\frac{d^4q}{(2\pi)^4}[ig_{\Upsilon(5S)BB}\epsilon^\mu_{\Upsilon(5S)}
(ip_{2\mu}-ip_{1\mu})]\nonumber\\&&\times[ig_{\Upsilon(nS)BB}
\epsilon^{\rho}_{\Upsilon(nS)}(-ip_{1\rho}-iq_{\rho})][g_{BBS}]\nonumber\\
&&\times\frac{1}{p_1^2-m_B^2}\frac{1}{p_2^2-m_B^2}\frac{1}{q^2-m_B^2}\mathcal{F}(q^2),\\
\nonumber\mathcal{M}_b&=&(i)^3\int\frac{d^4q}{(2\pi)^4}[-g_{\Upsilon(5S)BB}\varepsilon_{\mu\nu\alpha\beta}(-ip^{\mu}_{0})\epsilon^{\nu}_{\Upsilon(5S)}(ip^\alpha_{2})]
\nonumber\\&&\times[-g_{\Upsilon(nS)BB}\varepsilon_{\delta\tau\theta\phi}(ip^\delta_{3})\epsilon^\tau_{\Upsilon(nS)}(iq^{\theta})][-g_{B^*B^*S}]\nonumber
\\&&\times\frac{1}{p_1^2-m_B^2}\frac{-g^{\beta\rho}+p^\beta_2p^\rho_2/m_{B^*}^{2}}{p^2_2-m_{B^*}^{2}}\frac{-g^{\phi\rho}+q^\phi q^\rho/m_{B^*}^{2}}{q^2_2-m_{B^*}^{2}}
\mathcal{F}(q^2),\nonumber\\\\
\nonumber\mathcal{M}_c&=&(i)^3\int\frac{d^4q}{(2\pi)^4}[-g_{\Upsilon(5S)B^*B}\varepsilon_{\mu\nu\alpha\beta}(-ip^{\mu}_{0})\epsilon^{\nu}_{\Upsilon(5S)}(ip^\alpha_{1})]
\nonumber\\&&\times[-g_{\Upsilon(nS)B^{*}B}\varepsilon_{\delta\tau\theta\phi}(ip^\delta_{3})\epsilon^\tau_{\Upsilon(nS)}(-ip^{\theta}_{1})][g_{BBS}]
\nonumber\\&&\times\frac{-g^{\beta\phi}+p^\beta_1p^\phi_1/m_{B^*}^{2}}{p^2_1-m_{B^*}^{2}}\frac{1}{p^2_2-m_{B}^{2}}
\frac{1}{q^2-m_B^2}\mathcal{F}(q^2),\\
\nonumber\mathcal{M}_d&=&(i)^3\int\frac{d^4q}{(2\pi)^4}[-ig_{\Upsilon(5S)B^*B^*}\epsilon^{\mu}_{\Upsilon(5S)}((ip_{2\mu}-ip_{1\mu}
)g_{\nu\rho}\nonumber\\&&+(-ip_{0\rho}-ip_{2\rho})g_{\mu\nu}+(ip_{1\nu}+ip_{0\nu})g_{\mu\rho})]\nonumber\\
&&\times[-ig_{\Upsilon(nS)B^*B^*}\epsilon^{\phi}_{\Upsilon(nS)}((-ip_{1\phi}-iq_{\phi}
)g_{\alpha\beta}\nonumber\\&&+(ip_{3\beta}+ip_{1\beta})g_{\alpha\phi}+(iq_{\alpha}-ip_{3\alpha})g_{\beta\phi})]
[-g_{B^*B^*S}]\nonumber\\
&&\times\frac{-g^{\rho\alpha}+p^\rho_1p^\alpha_1/m_{B^*}^{2}}{p^2_1-m_{B^*}^{2}}
\frac{-g^{\nu\tau}+p^\nu_2p^\tau_2/m_{B^*}^{2}}{p^2_2-m_{B^*}^{2}}\nonumber\\&&\times
\frac{-g^{\beta\tau}+q^\beta q^\tau/m_{B^*}^{2}}{q^2-m_{B^*}^{2}}
\mathcal{F}(q^2),
\end{eqnarray}

\begin{eqnarray}
\nonumber\mathcal{M}_e&=&(i)^3\int\frac{d^4q}{(2\pi)^4}[-g_{\Upsilon(5S)BB}\varepsilon_{\mu\nu\alpha\beta}(-ip^{\mu}_{0})\epsilon^{\nu}_{\Upsilon(5S)}(ip^\alpha_{2})]
\nonumber\\&&\times[-g_{\Upsilon(nS)BB}\varepsilon_{\delta\tau\theta\phi}(ip^\delta_{3})\epsilon^\tau_{\Upsilon(nS)}(iq^{\theta})]
\nonumber\\&&\times[g_{f_2B^*B^*}\epsilon^{\rho\lambda}_{f_2}(g_{\rho\kappa}g_{\lambda\gamma}+g_{\rho\gamma}g_{\lambda\kappa}-g_{\rho\lambda}g_{\gamma\kappa})]\nonumber\\
&&\times\frac{1}{p_1^2-m_B^2}\frac{-g^{\beta\rho}+p^\beta_2p^\rho_2/m_{B^*}^{2}}{p^2_2-m_{B^*}^{2}}\frac{-g^{\phi\rho}+q^\phi q^\rho/m_{B^*}^{2}}{q^2-m_{B^*}^{2}}
\mathcal{F}(q^2),\nonumber\\\\
\nonumber\mathcal{M}_f&=&(i)^3\int\frac{d^4q}{(2\pi)^4}[-ig_{\Upsilon(5S)B^*B^*}\epsilon^{\mu}_{\Upsilon(5S)}((ip_{2\mu}-ip_{1\mu}
g_{\nu\rho})\nonumber\\&&+(-ip_{0\rho}-ip_{2\rho}g_{\mu\nu})+(ip_{1\nu}+ip_{0\nu}g_{\mu\rho}))]\nonumber\\
&&\times[-ig_{\Upsilon(nS)B^*B^*}\epsilon^{\phi}_{\Upsilon(nS)}((-ip_{1\phi}-iq_{\phi}
)g_{\alpha\beta}\nonumber\\&&+(ip_{3\beta}+ip_{1\beta})g_{\alpha\phi}+(iq_{\alpha}-ip_{3\alpha})g_{\beta\phi})]\nonumber\\
&&\times[g_{f_2B^*B^*}\epsilon^{\alpha\beta}_{f_2}(g_{\alpha\kappa}g_{\beta\gamma}+g_{\alpha\gamma}g_{\beta\kappa}-g_{\alpha\beta}g_{\kappa\gamma})]\nonumber\\
&&\times\frac{-g^{\rho\alpha}+p^\rho_1p^\alpha_1/m_{B^*}^{2}}{p^2_1-m_{B^*}^{2}}\frac{-g^{\nu\tau}+p^\nu_2p^\tau_2/m_{B^*}^{2}}{p^2_2-m_{B^*}^{2}}
\nonumber\\&&\times\frac{-g^{\beta\tau}+q^\beta q^\tau/m_B^{*2}}{q^2-m_{B^*}^{2}}
\mathcal{F}(q^2).
\end{eqnarray}

With above preparation, the amplitudes of $\Upsilon(5S)\to \Upsilon(nS)\pi^+\pi^-$ via the re-scattering can be parameterized as
\begin{eqnarray}
&&\mathcal{M}[\Upsilon(5S)\to
B^{(*)}\bar{B}^{(*)}\to\Upsilon(nS)(p_1)\pi^+(p_2)\pi^-(p_3)]_S\nonumber\\&&=
\big\{g^{(n)}_{0S} g_{\mu\nu}p_1\cdot
q+g_{0D}^{(n)}p_{1\mu}q_\nu\big\}\frac{\epsilon_{\Upsilon(5S)}^\mu
\epsilon_{\Upsilon(nS)}^{*\nu} g_{_{S
\pi\pi}} p_2 \cdot p_3}{q^2-m_S^2+i m_S \Gamma_S } ,
\label{1}\\
&&\mathcal{M}[\Upsilon(5S)\to B^{(*)}\bar{B}^{(*)}\to
\Upsilon(nS)(p_1)\pi^+(p_2)\pi^-(p_3)]_{f_2(1270)}
\nonumber\\&&=\big\{g_{2S}^{(n)} [g_{\mu \rho} g_{\nu
\lambda}+g_{\mu \lambda} g_{\nu \rho}] (p_1 \cdot q)^2 +
[g_{2D_1}^{(n)} g_{\mu \nu}
p_{1 \rho} q_{ \lambda}  \nonumber\\
&&\quad+ g_{2D_2}^{(n)} (g_{\mu \rho} q_{\nu} p_{1\lambda} +g_{\mu
\lambda} q_{\nu} p_{1\rho}) + g_{2D_3}^{(n)} (g_{\nu \lambda}
q_{\mu} p_{1\rho} \nonumber\\
&&\quad+g_{\nu \rho} q_{\mu} p_{1\lambda}) ]p_1 \cdot q
+g_{2G}^{(n)}q_{\mu} q_{\nu} p_{1\rho} p_{1\lambda} \big\}
\frac{\epsilon_{\Upsilon(5S)}^\mu \epsilon_{\Upsilon(nS)}^{*\nu}
\mathcal{P}_{f_2}^{\rho \lambda \alpha \beta}(q)}{q^2 -m_{f_2}^2 + i
m_{f_2}
\Gamma_{f_2}}\nonumber\\
&&\quad\times g_{f_2 \pi \pi}p_{2 \alpha} p_{2 \beta} ,\label{2}
\end{eqnarray}
corresponding to the contributions from the intermediate scalar
states $S=\{\sigma(600),f_0(980)\}$ and the tensor meson
$f_2(1270)$. In the above equation, $\mathcal{P}_{f_2}^{\rho \lambda
\alpha \beta}(q)$ is defined as
\begin{eqnarray}
\mathcal{P}_{f_2}^{\rho \lambda \alpha \beta}(q) = \frac{1}{2}
(\tilde{g}^{\rho \alpha} \tilde{g}^{\lambda \beta} + \tilde{g}^{\rho
\beta} \tilde{g}^{\lambda \alpha})- \frac{1}{3} \tilde{g}^{\rho
\lambda} \tilde{g}^{\alpha \beta}\nonumber
\end{eqnarray}
with $\tilde{g}^{\alpha \beta}= g^{\alpha \beta} - q^\alpha
q^\beta/m_{f_2}^2$. Then the differential decay width reads as
\begin{eqnarray}
d\Gamma =\frac{1}{3}\frac{1}{(2 \pi)^3} \frac{1}{32
M_{\Upsilon(5S)}^3} \overline{|\mathcal{M}|_{\mathrm{total}}^2}
dm_{\Upsilon \pi}^2 dm_{ \pi\pi}^2,
\end{eqnarray}
where $m_{\Upsilon \pi}^2 = (p_1 + p_2)^2$ and $m_{\pi\pi}^2 =(p_2
+p_3)^2$. The factor $1/3$ comes from an average over the
polarizations of the initial $\Upsilon(5S)$ state and in Ref.
\cite{Ali:2009es}, this factor was missing.

For the re-scattering process, the effective Lagrangian for
coupling bottomonia to the
bottomed mesons is determined  based on the
heavy quark effective theory \cite{Colangelo:2003sa}. The coupling
constants for $\Upsilon(5S) B^{(*)} B^{(*)}$ are evaluated by fitting the
partial decay widths while for $\Upsilon(nS)B^{(*)} B^{(*)}$
$(n=1,2)$ and $\mathcal{S} B^{(*)} B^{(*)}$, the coupling constants
are directly taken from Ref. \cite{Meng:2007tk}. In the re-scattering
picture, for the $\Upsilon(5S) \to \Upsilon(1S) \pi^+ \pi^-$ process,
the tensor meson $f_2(1270)$ should be included. Compared to the $S$-wave
coupling $f_2 B^\ast B^\ast$, the $D$-wave couplings $f_2 B B$ and $f_2
B B^\ast$ are negligible due to the high partial-wave suppression.
The coupling between $f_2(1270)$ and $B^\ast B^\ast $ has not been obtained from any measured reaction channels
yet; thus, in present work, we treat this coupling constant as
a free parameter to be fixed later. The coupling constants between the scalar mesons
and the final $\pi^+ \pi^-$ are $g_{\sigma \pi \pi}=16.2
\,\mathrm{GeV}^{-1}$ and $g_{f_0 \pi \pi}=2.40\,
\mathrm{GeV}^{-1}$, which are determined by fitting the corresponding partial widths.

In our model, besides the phase angles $\phi_{R}^{(n)}$ between the
re-scattering processes and the direct two-pion emission, just as indicated above, two more parameters
$\mathcal{A}$ and $\kappa$ are introduced for accounting the contribution from the direct process as
shown in Eq. (\ref{direct}). For calculating the re-scattering amplitudes, a form
factor is employed to describe the off-shell effects of the
exchanged mesons. In the calculations, the form factor takes the
monopole form, i.e., $\mathcal{F}(q^2)=(\Lambda^2-m_E^2)/(q^2-m_E^2)$,
where $m_E$ is the mass of the exchanged $B^{(*)}$ meson in the $B^{(*)}\bar{B}^{(*)}\to
\Upsilon(nS)\mathcal{S},\Upsilon(nS)f_2$ transitions shown in Fig. \ref{scalar}, and $\Lambda$ is
usually reparameterized as $\Lambda=m_E+ \alpha \Lambda_{QCD}$.
It is worth pointing out that such an adoption has a certain arbitrariness,
but the value $\Lambda$, which manifests all the unknown information about the non-perturbative QCD effects and the
inner structure of the involved mesons, is determined by
fitting data of various reactions; thus, it is believed that its value must fall in a reasonable range.
Thus the arbitrariness is relatively alleviated. In
present work, for $\Upsilon(5S) \to \Upsilon(nS) \pi^+ \pi^-$
$(n=1,2)$, $\alpha=2$ is adopted. The coefficients of the relevant Lorentz structures
in Eqs. (\ref{1})-(\ref{2}) $g_{0S}^{(n)}$, $g_{0D}^{(n)}$,
$g_{2S}^{(n)}$, $g_{2D_i}^{(n)}$ ($i=1,2,3$) and $g_{2G}^{(n)}$  are determined by calculating the hadronic loops. 

\begin{table}
\centering %
\caption{The resonance parameters (in units of GeV) used in this work \cite{Nakamura:2010zzi,Aitala:2000xu}. \label{Tab-Input}}
\begin{tabular}{cccccc}
 \toprule[1pt]
$m_{\Upsilon(5S)}$ & $10.870$ & $m_{\sigma}$ & $0.526$ & $\Gamma_{\sigma}$ & $0.302$ \\
 \midrule[1pt]
$m_{\Upsilon(1S)}$ & $9.460$ & $m_{f_0(980)}$ & $0.980$ & $\Gamma_{f_0(980)}$ & $0.070$ \\
$m_{\Upsilon(2S)}$ & $10.024$ & $m_{f_2}$ & $1.275$ & $\Gamma_{f_2}$ & $0.185$ \\
 \bottomrule[1pt]
\end{tabular}
\end{table}

\begin{figure}[htb]
\centering
\includegraphics[bb=600 390 20 520,scale=0.70,angle=0]{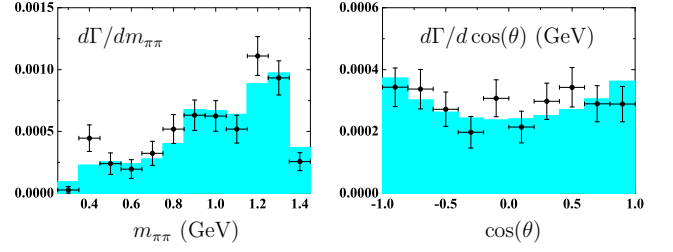}
\caption{(color online). Dipion invariant mass ($m_{\pi^+\pi^-}$)
distribution (left-hand side) and the $\cos \theta$ distribution (right-hand side) measured by Belle \cite{Abe:2007tk} for the final state $\Upsilon(1S) \pi^{+}
\pi^{-}$ (dots with error bars). The histograms are the best fit from
our model.\label{Fig-Fit5-1}}
\end{figure}

The $m_{\pi^+\pi^-}$ and $\cos \theta$ distributions
measured by the Belle Collaboration as well as the partial decay
widths $\Gamma_{\Upsilon(5S) \to \Upsilon(1S) \pi^+ \pi^-}= 0.59 \pm
0.04 \pm 0.09 $ MeV and $\Gamma_{\Upsilon(5S) \to \Upsilon(2S) \pi^+
\pi^-}= 0.89 \pm 0.07 \pm 0.16 $ MeV, are taken as inputs to our work. All other
input parameters, including the masses and widths of the involved
particles, are listed in Table \ref{Tab-Input}.
With the help of the MINUIT package,
we fit the Belle data of $\Upsilon(5S)\to\Upsilon(1S,2S)\pi^+\pi^-$ with the corresponding
parameters being fixed and listed in Tables \ref{Tab-Para1} and \ref{Tab-Para2}.

\begin{table}
\centering%
\caption{The parameters for $\Upsilon(5S) \to
\Upsilon(1S) \pi^+ \pi^-$ that are gained by fitting the Belle data. Here $g_{f_2} = g_{f_2 B^\ast B^\ast}
g_{f_2 \pi \pi}$.\label{Tab-Para1}}
\begin{tabular}{cccc}
 \toprule[1pt]
Parameter  & Value  & Parameter & Value (Rad)\\
 \midrule[1pt]
$F$      & $0.186 \pm 0.061$    & $\phi_{\sigma(600)}^{(1)}$ &  $- 2.638\pm0.735$    \\
$\kappa$  & $0.459 \pm 0.084$    & $\phi_{f_0(980)}^{(1)}$ &  $1.539\pm0.741$    \\
$g_{f_2}$& $12.361 \pm 20.109$  & $\phi_{f_2(1270)}^{(1)}$ &  $ -1.028\pm2.050$    \\
 \bottomrule[1pt]
\end{tabular}
\end{table}

\begin{table}
\centering%
\caption{The fitted parameters for $\Upsilon(5S) \to \Upsilon(2S)
\pi^+ \pi^-$.\label{Tab-Para2}}
\begin{tabular}{cccc}
 \toprule[1pt]
Parameter  & Value  & Parameter & Value (Rad)\\
 \midrule[1pt]
$F$      & $2.315 \pm 1.904$    & $\phi_{\sigma(600)}^{(2)}$ &  $-0.297\pm0.567$    \\
$\kappa$  & $0.572 \pm 0.283$    & $\phi_{f_0(980)}^{(2)}$ &  $-3.140\pm4.532$    \\
 \bottomrule[1pt]
\end{tabular}
\end{table}
\begin{figure}
\includegraphics[bb=600 390 20 520,scale=0.70,angle=0]{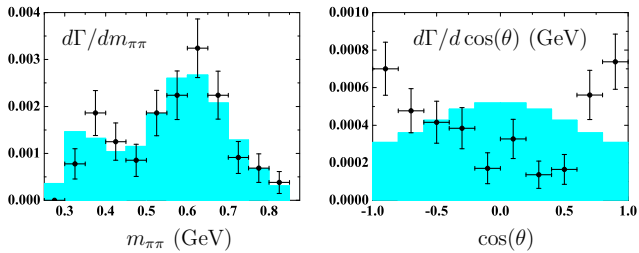}
\caption{(color online). The comparison between the fitting result (histogram) for $\Upsilon(5S) \to \Upsilon(2S) \pi^+ \pi^-$ and the Belle data (dots with error bars) \cite{Abe:2007tk}. \label{Fig-Fit5-2}}
\end{figure}

The dipion invariant mass distribution $d\Gamma/dm_{\pi^+\pi^-}$ and the
angular distribution $d\Gamma/d\cos \theta $ measured by the Belle
Collaboration for $\Upsilon(5S) \to \Upsilon(1S) \pi^+ \pi^-$ are
shown in Fig. \ref{Fig-Fit5-1}. The shaded histograms are the
corresponding theoretical prediction by our model. The
parameters for $\Upsilon(5S) \to \Upsilon(1S) \pi^+ \pi^-$ are
listed in Table \ref{Tab-Para1}, yielding an integrated decay width
of $\Gamma_{\Upsilon(5S) \to \Upsilon(1S) \pi^+ \pi^-}=0.54 $ MeV.
The consistency between our results and the Belle data indicates that
our model can naturally describe the anomalous production rate of $\Upsilon(1S)\pi^+\pi^-$
near the peak of $\Upsilon(5S)$ well, and, moreover, the predicted dipion invariant mass distribution and the $\cos\theta$
distribution also coincide with the data.

For $\Upsilon(5S) \to \Upsilon(2S) \pi^+ \pi^-$, we carry out a similar calculation. The
shaded histograms are our best fit to the experimental data, and
the corresponding parameters are listed in Table \ref{Tab-Para2}. The
integrated decay width of $\Gamma_{\Upsilon(5S) \to \Upsilon(2S)
\pi^+ \pi^-}=0.845 $ MeV. Our results also confirm that the contribution from $f_0(980)$
is rather small compared to the $\Upsilon(5S)\to \Upsilon(2S)\pi^+\pi^-$ process.

However, one notices a discrepancy. As shown in Fig. \ref{Fig-Fit5-2}, the dipion invariant mass distribution of the
$\Upsilon(2S)\pi^+\pi^-$ production near the peak of $\Upsilon(5S)$
is well reproduced by our model. However, applying the same fitting parameters, the predicted
$d\Gamma/d\cos\theta$ of $\Upsilon(5S)\to\Upsilon(2S)\pi^+\pi^-$ (the histogram on the left panel of Fig. \ref{Fig-Fit5-2}),
displays a different behavior from the Belle data for the $\Upsilon(2S)\pi^+\pi^-$ channel
(dots with error bars in the right-hand diagram of  Fig. \ref{Fig-Fit5-2}).

In summary, stimulated by the recent Belle observation of anomalously large
production rates of $\Upsilon(1S,2S)\pi^+\pi^-$  near the peak of $\Upsilon(10860)$, carefully studying the previous works along the line,
we suggest that both the direct
dipion emission process and the processes via intermediate physical states which are the so-called
final state interactions, contribute to the amplitude, and their interference results in the observed dipion
emission of $\Upsilon(10860)$. In our scenario, the inverse rates $\Gamma(\Upsilon(10860)\to \Upsilon(2S)\pi^+\pi^-)>\Gamma(\Upsilon(10860)\to \Upsilon(1S)\pi^+\pi^-)$
can also be naturally understood, i.e., their rates are determined by the interference between the
contributions of the direct emission and the final interactions. 

Fitting the decay rates of $\Upsilon(10860)\to \Upsilon(1S,2S)\pi^+\pi^-$, we further theoretically
investigate the dipion invariant mass and the $\cos\theta$ distributions of $\Upsilon(10860)\to \Upsilon(1S,2S)\pi^+\pi^-$ and make a comparison with the Belle data.
Indeed, it is observed that if the final state interactions overwhelmingly dominate the transitions
$\Upsilon(10860)\to \Upsilon(1S)+\pi^+\pi^-$ and $\Upsilon(10860)\to \Upsilon(2S)\pi^+\pi^-$, the lineshapes of the differential widths over the dipion invariant
mass and $\cos\theta$ cannot be well fitted. It indicates that the interference effect plays a crucial role for fully
understanding the Belle observation \cite{Abe:2007tk}.

What is more important is that our model demonstrated in this paper shows that the tetraquark scenario proposed by Ali {\it et al}.
\cite{Ali:2009es,Ali:2009pi,Ali:2010pq} does not provide a satisfactory  understanding of the anomalous
$e^+e^-\to \psi(2S)\pi^+\pi^-$ production at $\sqrt{s}=10.870$ GeV. However, one cannot rule out
that there might be a fraction of the tetraquark component in $\Upsilon(10860)$ that also contributes to the dipion transition. But, so far, it seems that a contribution from such
an exotic state is not necessary for just understanding the Belle data.
Instead, our study presented in this paper indicates that the Belle observation can be naturally explained by the interference between the direct dipion emission and the
final state interactions.

It is worth pointing out that in our model, our theoretical predictions on both anomalous production rates of $\Gamma(\Upsilon(10860)\to \Upsilon(2S)\pi^+\pi^-)$
and $\Gamma(\Upsilon(10860)\to \Upsilon(1S)\pi^+\pi^-)$ coincide with the data of the Belle collaboration, and also satisfactorily describe
the dipion invariant mass and the $\cos\theta$ distributions of $\Upsilon(10860)\to \Upsilon(1S)\pi^+\pi^-$ as well as the dipion invariant mass distribution of
$\Upsilon(10860)\to \Upsilon(2S)\pi^+\pi^-$. However, a new and intriguing puzzle is proposed since the predicted $d\Gamma/d\cos\theta$ of $\Upsilon(10860)\to \Upsilon(2S)\pi^+\pi^-$ is inverse to the Belle data just presented in the right panels of Figs. \ref{Fig-Fit-ali-1} and \ref{Fig-Fit5-2}. Associated with further theoretical exploration, future experimental study from Belle-II and SuperB will be helpful to clarify this new puzzle and give a definite conclusion.

\noindent {\it Acknowledgement}: We would like to thank C. Hambrock, Wei Wang and Feng-Kun Guo for useful communication. This project is supported by the
National Natural Science Foundation of China under Grants No.
11175073, No. 10905077, No. 11005129, No. 11035006, No. 11047606; the Ministry of Education of China (FANEDD under Grant
No. 200924, DPFIHE under Grant No. 20090211120029, NCET under
Grant No. NCET-10-0442, the Fundamental Research Funds for the
Central Universities); and the West Doctoral Project of the Chinese
Academy of Sciences.

\end{document}